\documentclass[%
  twocolumn, 
  floatfix,
  aps, 
  prl, 
  final, letterpaper, 
  amsmath, amsfonts, amssymb
]{revtex4-2}

\usepackage{graphicx}% Include figure files
\usepackage{hyperref}% add hypertext capabilities
\usepackage[utf8]{inputenc}

\usepackage{physics}
\usepackage{xcolor}
\begin{document}

\title{Herd Immunity and Epidemic Size in Networks with Vaccination Homophily}

\author{Takayuki Hiraoka}
%\email{takayuki.hiraoka@aalto.fi}
\author{Abbas K.~Rizi}
\author{Mikko Kivel\"{a}}
\author{Jari Saram\"{a}ki}
\affiliation{Department of Computer Science, Aalto University, 00076 Espoo, Finland} 

% \date{\today}

\begin{abstract}
We study how the herd immunity threshold and the expected epidemic size depend on homophily with respect to vaccine adoption. We find that the presence of homophily considerably increases the critical vaccine coverage needed for herd immunity and that strong homophily can push the threshold entirely out of reach. The epidemic size monotonically increases as a function of homophily strength for a perfect vaccine, while it is maximized at a nontrivial level of homophily when the vaccine efficacy is limited. Our results highlight the importance of vaccination homophily in epidemic modeling.
\end{abstract}

\maketitle

\paragraph{Introduction.}
In the paradigmatic Susceptible-Infectious-Recovered (SIR) model of infectious disease in a fully mixed population~\cite{AM,Hethcote2000Mathematics}, so-called herd immunity is reached when the fraction $\pi_\mathrm{v}$ of the population that is immune to the disease through vaccination or previous infection is larger than
\begin{equation}
   \pi_\mathrm{v}^\mathrm{c} = 1-\frac{1}{R_0}\,,
   \label{eq:trad}
\end{equation}
where $R_0$ denotes the basic reproduction number, i.e., the expected number of secondary cases produced by a typical infectious individual in a fully susceptible population.
Here, herd immunity means that the disease cannot spread in the population because each infected individual can only transmit the infection to less than one other individual on average; that is, the effective reproduction number $R_\mathrm{eff}=\left(1-\pi_\mathrm{v}\right)R_0<1$. Consequently, not only those who are vaccinated but also the unvaccinated individuals are collectively protected from the disease.

This model assumes homogeneous mixing where individuals interact with each other randomly and independently of their properties such as their vaccination status. However, this is a premise that may be too simplistic for modeling real-world populations, which often exhibit inhomogeneous mixing patterns that can lead to nontrivial epidemic outcomes~\cite{Newman2003Mixing, Lloyd-Smith2005Superspreading, Mossong2008Social, Hebert-Dufresne2020Beyond}. One of the inhomogeneities that would be particularly relevant to vaccine-induced herd immunity is the correlation between the vaccination status of interacting individuals~\cite{Salathe2008Effect, Mbah2012Impact, Barclay2014Positive, Edge2015Seasonal, Truelove2019Characterizing, Kadelka2021Effect}. When this correlation exists, the vaccinated and unvaccinated individuals have different compositions of vaccinated and unvaccinated neighbors. Let us introduce the term \emph{vaccination homophily} to represent mixing patterns that are assortative with respect to vaccination status, so that connections are more probable within the vaccinated and unvaccinated populations than between them. In this Letter, we investigate the effect of vaccination homophily on the herd immunity threshold and the expected epidemic size.

\paragraph{Model.}
To this end, we formulate a random network theory of epidemic spreading under homophily with respect to the adoption of an immunity-inducing vaccine. The links in the network represent transmissible contacts between individuals, i.e., a susceptible individual will get infected if connected to an infected individual. We refer to this network as the \emph{transmission network} to avoid confusion with the \emph{contact network}. Each link in the contact network will let the disease be transmitted through it with a certain probability; the links on which transmission actually takes place constitute the transmission network~\cite{Newman2002Spread, Pastor-Satorras2015Epidemic}. Here, we do not explicitly consider this probabilistic transmission process, but rather take the transmission network as a given. 

Within the population, a fraction $\pi_\mathrm{v}$ of the population adopts the vaccine, while the remaining fraction $\pi_\mathrm{u} = 1 - \pi_\mathrm{v}$ is not vaccinated. Vaccination homophily can be expressed in terms of the bias in the probabilities of connections within the two groups. Let us denote the conditional probability that a random neighbor of an individual is vaccinated given that the individual is vaccinated by $\pi_\mathrm{vv}$ and, similarly, the conditional probability that a random neighbor of an unvaccinated individual is not vaccinated by $\pi_\mathrm{uu}$. Assuming that the average degrees (numbers of connections) of the vaccinated and unvaccinated populations are equal, the two probabilities are related as $\pi_\mathrm{uu} = 1 - (1 - \pi_\mathrm{vv}) \pi_\mathrm{v} / \pi_\mathrm{u}$.

The problem of using the connection probabilities $\pi_\mathrm{vv}$ and $\pi_\mathrm{uu}$ as measures of homophily is that they are not ``orthogonal" to  $\pi_\mathrm{v}$ so that even if we fix the value of $\pi_\mathrm{vv}$, the strength of homophily varies with different values of $\pi_\mathrm{v}$. Moreover, the two connection probabilities are coupled in a nonlinear manner, making it difficult to justify using either of them as a representative measure of the homophily of the entire network structure. To address these issues, we adopt the Coleman homophily index, originally proposed for social network analysis~\cite{Coleman1958Relational} and defined by
\begin{equation}
    h = \frac{\pi_\mathrm{vv} - \pi_\mathrm{v}}{1 - \pi_\mathrm{v}} = \frac{\pi_\mathrm{uu} - \pi_\mathrm{u}}{1 - \pi_\mathrm{u}}\,.
\end{equation}
This measure has desirable axiomatic properties: i) it is an increasing function of both $\pi_\mathrm{vv}$ and $\pi_\mathrm{uu}$, ii) 
it is symmetric for the vaccinated and unvaccinated populations, and iii)
it takes a value of zero when the mixing is homogeneous (no homophily) and a value of one when all links are inside the two groups, that is, 
$\pi_\mathrm{vv} = \pi_\mathrm{uu} = 1$. A negative value implies that the network is heterophilic in terms of vaccination status. Note that the connection probabilities 
$\pi_\mathrm{vv} = \pi_\mathrm{v} + \pi_\mathrm{u} h$ 
and
$\pi_\mathrm{uu} = \pi_\mathrm{u} + \pi_\mathrm{v} h$ 
must be positive and therefore the Coleman homophily index is bounded from below as  
$h \geq \max \left( - \pi_\mathrm{v} / \pi_\mathrm{u}, - \pi_\mathrm{u} / \pi_\mathrm{v} \right)$.

We consider the transmission network structure where $\pi_\mathrm{v}$, $h$, and the degree distribution $P(k)$ are specified but otherwise maximally randomized. Neglecting the rare cycles, we can identify the basic reproduction number as the mean excess degree of the network, i.e., the expected number of other neighbors that a randomly chosen neighbor of a randomly chosen node has, as $R_0 = \left \langle k^2 \right \rangle/\left \langle k \right \rangle - 1$~\cite{Molloy1995Critical, Newman2001Random, Trapman2016Inferring}.

We consider a class of epidemic models where infection induces complete and permanent immunity, whereas the immunity induced by vaccines is generally incomplete. There are two effects of vaccine protection that are of interest for modeling herd immunity~\cite{Halloran1999Design, Farrington2003Vaccine}. First, the vaccine can reduce the probability that the recipient becomes infected upon exposure. This reduction is referred to as the efficacy against susceptibility and denoted by $f_\mathrm{S}$~\footnote{
Here, we assume that the vaccine induces full immunity to a fraction $f_\mathrm{S}$ of the vaccinated individuals, but leaves the remainder fully susceptible. In this model, the vaccine is \emph{all-or-nothing}, in contrast to \emph{leaky} vaccines which reduce the susceptibility of every recipient by an equal degree. The two models are equivalent under the assumption that the network is locally tree-like, on which our study is based.}. 
Second, individuals who are infected despite being vaccinated may have a lower probability of transmitting the infection to others. We represent this with the efficacy against infectiousness, $f_\mathrm{I}$, defined as the reduction in the secondary infection rate.

Under this setup, the herd immunity threshold and the expected final size of a large epidemic can be derived from the structure of the transmission network alone, without explicitly considering the epidemic dynamics. In the following, we leverage the theory of branching processes and percolation theory to investigate these quantities of interest.

\paragraph{Herd immunity threshold.}

\begin{figure}
    \centering
    \includegraphics[width=\columnwidth]{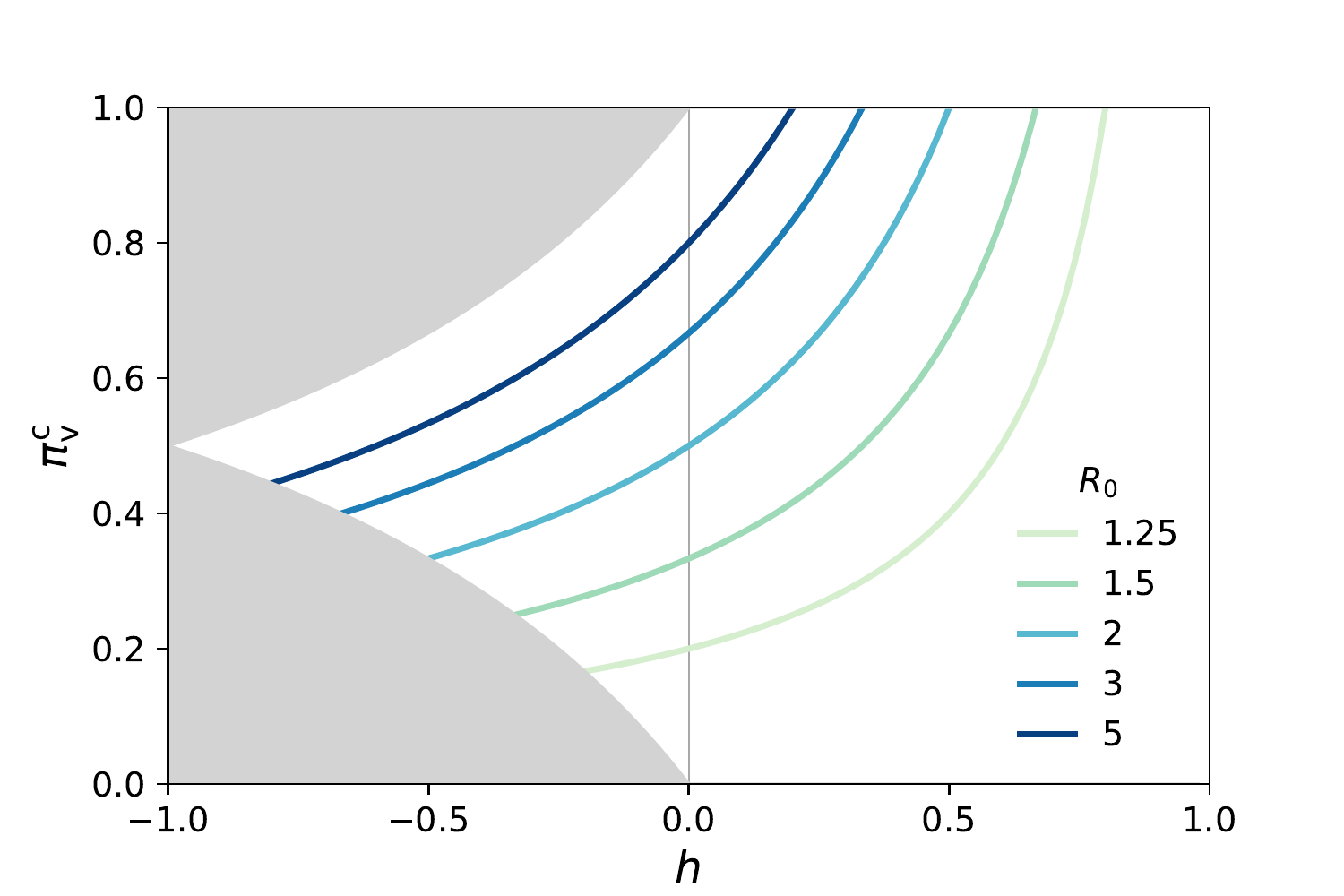}
    \caption{Critical coverage $\pi_\mathrm{v}^\mathrm{c}$ of a perfect vaccine required for herd immunity as a function of homophily strength $h$ for different values of basic reproduction number $R_0$. Positive and negative values of $h$ imply homophily and heterophily, respectively. The area shaded in gray represents the parameter region where the network is unrealizable.}
    \label{fig:critical_vax_rate}
\end{figure}

For a heterogeneous population consisting of multiple subpopulations, we can use the next-generation matrix (NGM) method~\cite{Diekmann1990Definition, Diekmann2012Mathematical} to identify the vaccination threshold $\pi_\mathrm{v}^\mathrm{c}$ above which the disease cannot spread. While the NGM method was originally developed for epidemic dynamics described by ordinary differential equations, it can be naturally interpreted as a description of the local structure of the transmission network by a multi-type branching process where the branching factor is the excess degree of the network. Let us denote by $I_\mathrm{v}^{(m)}$ and $I_\mathrm{u}^{(m)}$ the number of infections in the vaccinated and unvaccinated populations, respectively, at generation $m$ from an index case (the first infected individual). Assuming a locally tree-like network, we can write the following recurrence equations under a mean-field approximation:
\begin{gather}
    I_\mathrm{v}^{(m+1)} = (1 - f_\mathrm{S}) R_0 [ (1 - f_\mathrm{I}) \pi_\mathrm{vv} I_\mathrm{v}^{(m)} + \pi_\mathrm{uv} I_\mathrm{u}^{(m)} ], \label{eq:branching_general_v}\\
    I_\mathrm{u}^{(m+1)} = R_0 [ (1 - f_\mathrm{I}) \pi_\mathrm{vu} I_\mathrm{v}^{(m)} + \pi_\mathrm{uu} I_\mathrm{u}^{(m)} ],
    \label{eq:branching_general_u}
\end{gather}
where $\pi_\mathrm{uv} = 1 - \pi_\mathrm{uu}$ and $\pi_\mathrm{vu} = 1 - \pi_\mathrm{vv}$ are the conditional probabilities that a link from one group points to the other. 
By writing $\mathbf{I}^{(m+1)}=\mathbf{A} \mathbf{I}^{(m)}$, where $\mathbf{I}^{(m)}=(I_\mathrm{v}^{(m)},I_\mathrm{u}^{(m)})^\intercal$ and 
\begin{equation*}
\mathbf{A} = R_0 
\begin{pmatrix} 
(1 - f_\mathrm{S}) (1 - f_\mathrm{I}) \pi_\mathrm{vv} & (1 - f_\mathrm{S}) \pi_\mathrm{uv}\\
(1 - f_\mathrm{I}) \pi_\mathrm{vu} & \pi_\mathrm{uu}
\end{pmatrix},
\end{equation*}
we see that the infection eventually dies out after a finite number of generations if all the eigenvalues of the next-generation matrix $\mathbf{A}$ have an absolute value of less than one. That is, at the critical point, the spectral radius $\rho(\mathbf{A})=1$.

By reparameterizing the connection probabilities with $\pi_\mathrm{v}$ and $h$, the critical vaccine coverage needed for herd immunity is given by
\begin{equation}
    \pi_\mathrm{v}^\mathrm{c} = \frac{1 - \epsilon R_0 h}{(1 - \epsilon) (1 - h)}\left( 1 - \frac{1}{R_0} \right),
    \label{eq:critial_vax_rate_general}
\end{equation}
where we define $\epsilon = (1 - f_\mathrm{S}) (1 - f_\mathrm{I})$ and require $\epsilon \leq 1 / R_0$. For $\epsilon > 1/R_0$, the vaccination threshold disappears and herd immunity becomes unattainable.
For a perfect vaccine with $f_\mathrm{S} = 1$ and/or $f_\mathrm{I} = 1$, we have
\begin{equation}
    \pi_\mathrm{v}^\mathrm{c} = \frac{1}{1 - h} \left(1 - \frac{1}{R_0}\right),
    \label{eq:full-piv}
\end{equation}
which reduces to the well-known threshold of Eq.~(\ref{eq:trad}) for homogeneous mixing with $h=0$. 

Equation~(\ref{eq:full-piv}) indicates that if the homophily strength $h$ increases, so does the vaccine coverage $\pi_\mathrm{v}^\mathrm{c}$ required for herd immunity (see Fig.~\ref{fig:critical_vax_rate}). In other words, the presence of homophily makes herd immunity harder to reach.
Notably, the threshold occurs at $\pi_\mathrm{v}^\mathrm{c}= 1$ for \begin{equation}
    h \geq \frac{1}{R_0}\,,
\end{equation}
implying that above this critical strength of homophily, one cannot attain herd immunity at all unless the entire population is vaccinated. That is, no matter how small the unvaccinated population is, there will always be a nonzero probability of a large epidemic within this population. 

Finally, we note that the above discussion is independent of the specific shape of the degree distribution---the equations are valid for any degree distribution $P(k)$ with mean excess degree $R_0$.

\paragraph{Epidemic size.}

\begin{figure*}
    \centering
    \includegraphics[width=0.8\textwidth]{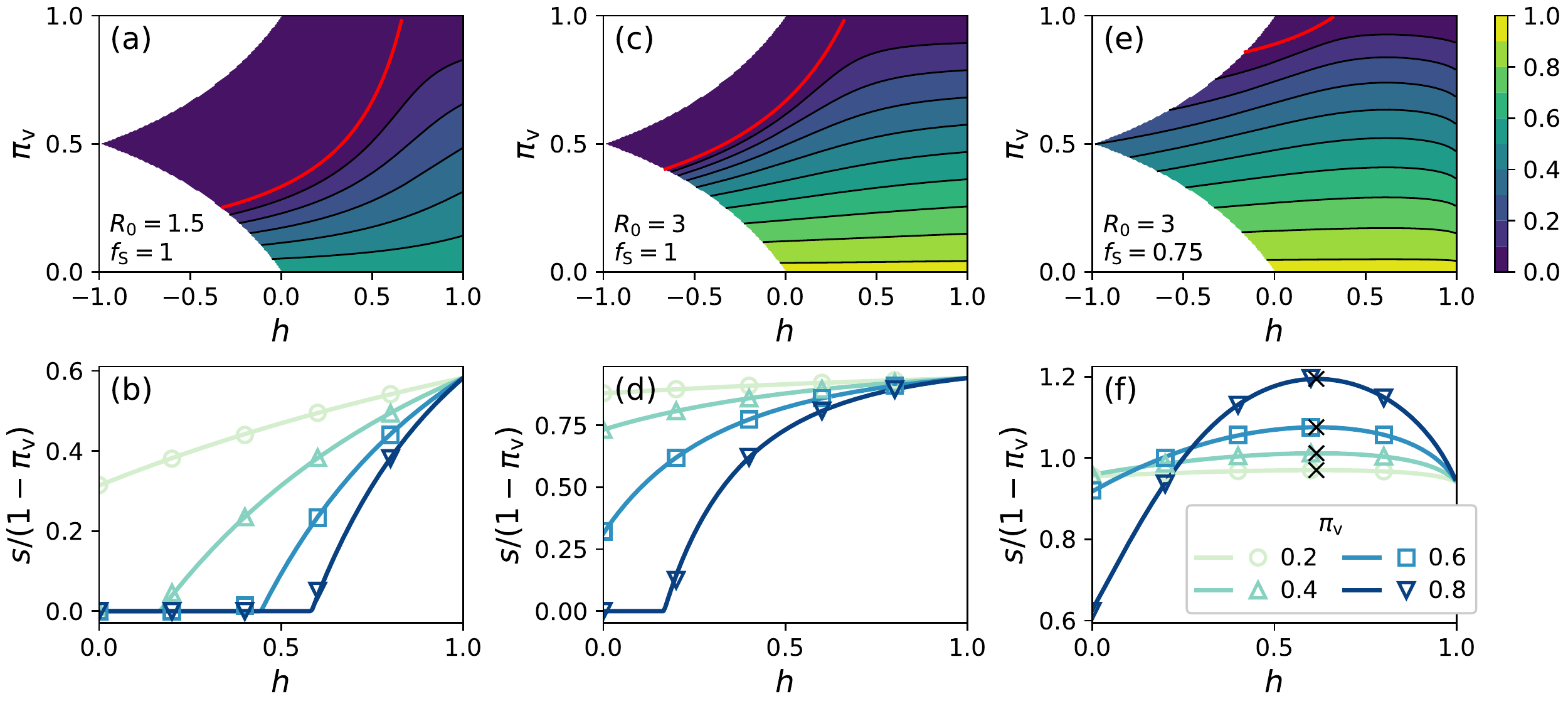}
    \caption{Epidemic size in Poisson networks as a function of homophily strength $h$ and vaccine coverage $\pi_\mathrm{v}$. Top row: Two-dimensional heatmaps representing the epidemic size. The solid red line in each panel denotes the vaccination threshold. We represent contours of the epidemic size at 0.1 intervals by different colors and solid black lines. Bottom row: Epidemic size divided by the size of the unvaccinated population. Theoretical predictions (in lines) are compared with the giant component sizes obtained by simulating networks of size $N = 10^5$ (in symbols). The details of the network simulation can be found in the Supplemental Material. (a) and (b) show the results for $R_0 = 1.5$ and a perfect vaccine, (c) and (d) are for $R_0 = 3$ and a perfect vaccine, and (e) and (f) are for $R_0 = 3$ and an imperfect vaccine with $f_\mathrm{S} = 0.75$. If the vaccine is perfect, only the unvaccinated individuals contract the disease; thus, the vertical axis in (b) and (d) corresponds to the fraction of the unvaccinated population that will be infected. The cross symbols in (f) indicate the maximum of each curve. Note that the homophily strength at which the epidemic size takes the maximum is independent of $\pi_\mathrm{v}$.}
    \label{fig:episizes_poisson_main}
\end{figure*}

When the vaccine coverage is below the threshold, an outbreak can result in an epidemic that infects a substantial fraction of the population. The size of such an epidemic coincides with the size of the giant component of the transmission network because all the individuals in a connected component will be infected if the index case belongs to the same component~\cite{Newman2001Random, Newman2002Spread}. Let us denote the probability that a link pointing to a vaccinated node does not lead to the giant component by $\phi_\mathrm{v}$ and the equivalent probability for an unvaccinated node by $\phi_\mathrm{u}$. These probabilities are subject to the following consistency equations:
\begin{align}
    \phi_\mathrm{v} &= 
    \begin{aligned}[t]
        f_\mathrm{S} + &(1 - f_\mathrm{S}) \\
        &\times g_1 \boldsymbol{(} f_\mathrm{I} + (1 - f_\mathrm{I}) (\pi_\mathrm{vv} \phi_\mathrm{v} + \pi_\mathrm{vu} \phi_\mathrm{u}) \boldsymbol{)},
    \end{aligned}
    \label{eq:consistency_v_general}\\
    \phi_\mathrm{u} &= g_1(\pi_\mathrm{uv} \phi_\mathrm{v} + \pi_\mathrm{uu} \phi_\mathrm{u}),
    \label{eq:consistency_u_general}
\end{align}
where $g_1(x) = \sum_{k=1}^\infty k P(k) x^{k-1} / \langle k \rangle$ denotes the probability generating function of excess degree. 
Having solved the above consistency equations for $\phi_\mathrm{v}$ and $\phi_\mathrm{u}$, we can compute the size of the vaccinated and unvaccinated populations contained in the giant component as
\begin{align}
    s_\mathrm{v} &= 
    \begin{aligned}[t]
        &(1 - f_\mathrm{S}) \pi_\mathrm{v} \\
        &\times [1 - g_0 \boldsymbol{(} f_\mathrm{I} + (1 - f_\mathrm{I}) (\pi_\mathrm{vv} \phi_\mathrm{v} + \pi_\mathrm{vu} \phi_\mathrm{u}) \boldsymbol{)}],
    \end{aligned}
    \label{eq:size_v_general}\\
    s_\mathrm{u} &= \pi_\mathrm{u} [ 1 - g_0(\pi_\mathrm{uv} \phi_\mathrm{v} + \pi_\mathrm{uu} \phi_\mathrm{u}) ],
    \label{eq:size_u_general}
\end{align}
respectively, where $g_0(x) = \sum_{k=0}^\infty P(k) x^k$ is the probability generating function of the degree distribution $P(k)$. The total size of the giant component is the sum of these two fractions $s= s_\mathrm{u} + s_\mathrm{v}$.

As an illustration, let us solve the above equations for a random network with a Poisson degree distribution $P(k) = {\langle k \rangle^k e^{-\langle k \rangle}}/{k!}$. For this network, the excess degree distribution is identical to the degree distribution, and hence $\langle k \rangle=R_0$. Given this degree distribution, we get $g_0(x) = g_1(x) = \exp[-R_0 (1 - x)]$.
In the thermodynamic limit and in the absence of homophily ($h=0$), this random network model with the Poisson degree distribution reduces to the Erd\H{o}s-R\'{e}nyi (ER) random graph ensemble, which is equivalent to homogeneous mixing. In other words, our model represents the simplest deviation from the ER model through the addition of homophily that biases the randomness of links.

First, let us consider the case of a perfect vaccine, for which $\phi_\mathrm{v} = 1$. Eq.~(\ref{eq:consistency_u_general}) now becomes 
\begin{equation}
    \phi_\mathrm{u} = \exp[- R_0 \pi_\mathrm{uu} (1 - \phi_\mathrm{u})],
\end{equation}
which has an analytical solution
\begin{equation}
    \phi_\mathrm{u} = - \frac{ W \boldsymbol{(} - R_0 \pi_\mathrm{uu} \,\exp(- R_0 \pi_\mathrm{uu}) \boldsymbol{)} }{R_0 \pi_\mathrm{uu}}\,.
    \label{eq:consistency_u_perfect_solution}
\end{equation}
Here, $W(\cdot)$ denotes the Lambert $W$-function. The giant component size is then calculated from Eq.~(\ref{eq:size_u_general}) as
\begin{equation}
    s = s_\mathrm{u} = \pi_\mathrm{u}\{1 - \exp[- R_0 \pi_\mathrm{uu} (1 - \phi_\mathrm{u})]\},
\label{eq:perfsize}
\end{equation}
where all infections are restricted to the unvaccinated population.

Figure~\ref{fig:episizes_poisson_main}(a--d) shows the solution of Eq.~(\ref{eq:perfsize}). The main observation is that the expected epidemic size always increases with homophily strength $h$. 
The difference in epidemic size under strong and weak homophily is especially significant when the vaccine coverage $\pi_\mathrm{v}$ is not small. As an example, for a disease with $R_0=1.5$, the homogeneous mixing assumption leads to the prediction that the vaccination threshold is 33\%. However, even if the vaccine coverage is well above this threshold, strong homophily can still let the disease spread in the unvaccinated population and infect up to 58\% of it (see Fig.~\ref{fig:episizes_poisson_main}(b)).

In the case of imperfect vaccines, the coupled consistency equations are not analytically tractable. The solution of Eq.~\eqref{eq:consistency_u_general} is given by
\begin{equation}
    \phi_\mathrm{u} = - \frac{ W \boldsymbol{(} - R_0 \pi_\mathrm{uu} \,\exp[ - R_0 [1 - (1 - \pi_\mathrm{uu}) \phi_\mathrm{v}] ] \boldsymbol{)} }{R_0 \pi_\mathrm{uu}}\,, 
    \label{eq:consistency_u_general_solution}
\end{equation}
whereas for $f_\mathrm{S} < 1$ and $f_\mathrm{I} < 1$, Eq.~\eqref{eq:consistency_v_general} leads to
\begin{equation}
    \phi_\mathrm{u} = \frac{1}{1 - \pi_\mathrm{vv}} \left( 1 - \pi_\mathrm{vv} \phi_\mathrm{v} + \frac{1}{(1 - f_\mathrm{I}) R_0} \,\log \frac{\phi_\mathrm{v} - f_\mathrm{S}}{1 - f_\mathrm{S}} \right).
    \label{eq:consistency_v_general_solution}
\end{equation}
We can numerically solve for $\phi_\mathrm{v}$ by equating the right hand sides of Eqs.~\eqref{eq:consistency_u_general_solution} and \eqref{eq:consistency_v_general_solution}. Plugging the results into Eqs.~\eqref{eq:size_v_general} and \eqref{eq:size_u_general} yields the giant component size. 

\begin{figure}
    \centering
    \includegraphics[width=\columnwidth]{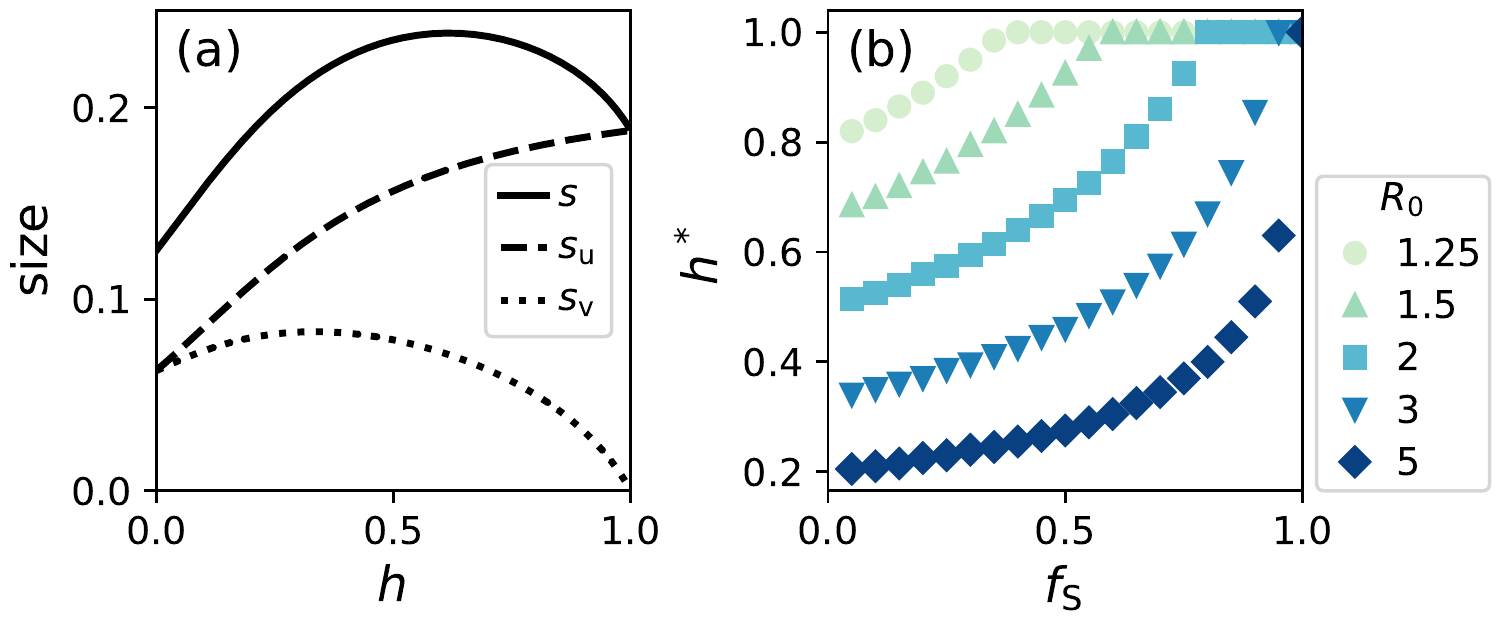}
    \caption{Effects of vaccination homophily for imperfect vaccines. (a) The sizes of vaccinated population $s_\mathrm{v}$ and unvaccinated population $s_\mathrm{u}$ in the epidemic of size $s$. The parameters are: $R_0 = 3,\, \pi_\mathrm{v} = 0.8,\, f_\mathrm{S} = 0.75$. (b) The homophily strength $h^*$ maximising $s$ as a function of $f_\mathrm{S}$ and $R_0$.}
    \label{fig:imperfect_vaccine}
\end{figure}

In what follows, we present the results for $f_\mathrm{I} = 0$ and only vary the efficacy against susceptibility, $f_\mathrm{S}$, for the sake of simplicity. Figure~\ref{fig:episizes_poisson_main}(e,~f) shows the epidemic size under the coverage of an imperfect vaccine. As expected, a smaller efficacy leads to a larger epidemic and a higher vaccination threshold. Unexpectedly, contrary to the case of perfect immunization, the epidemic size first grows and then shrinks with increasing homophily.
This can be attributed to the following competing mechanisms affected by increased levels of homophily:
(1) Similarly to the case of a perfect vaccine, more unvaccinated individuals will be infected as they are connected to fewer immune individuals and more densely within themselves, making them less protected by the herd immunity effect. 
(2) An imperfect vaccine leaves a part of the vaccinated population susceptible to breakthrough infections. In the weak homophily regime, more vaccinated individuals may contract the disease due to the larger epidemic in the unvaccinated population. The risk of breakthrough infection decreases as they become less connected with the unvaccinated population in the strong homophily regime.
Figure~\ref{fig:imperfect_vaccine}(a) gives an example of the two competing processes, where given $\pi_\mathrm{v}=0.8$, $R_0=3$, and vaccine efficacy $f_\mathrm{S}=0.75$, the final epidemic size varies between 13\% and 24\%, reaching its peak around $h=0.62$. 

As a consequence of the competition, the total number of infected individuals is maximized, in general, at a nontrivial level of homophily $h^*$, which depends on $f_\mathrm{S}$ and $R_0$ but not on the vaccine coverage $\pi_\mathrm{v}$. The smaller the $R_0$ and higher the value of $f_\mathrm{S}$, the higher the strength of homophily  $h^*$ that leads to the worst overall outcome (see Fig.~\ref{fig:imperfect_vaccine}(b)). 
In other words, a highly infectious disease countered by a vaccine with low efficacy spreads maximally in a population with a medium level of vaccination homophily, while less infectious diseases generally benefit from higher levels of homophily, especially if the vaccine efficacy is high. The maximum impact of homophily on epidemic size is further discussed in the Supplemental Material.

In the above discussion, we presented the results for the case where the transmission network has a Poisson degree distribution and the efficacy against infectiousness $f_\mathrm{I} = 0$. These conditions can be altered. In the Supplemental Material, we calculate the epidemic size for transmission networks with more realistically heterogeneous excess degrees that follow the negative binomial distribution. We also discuss the case where both $f_\mathrm{S}$ and $f_\mathrm{I}$ are varied. In both cases, the epidemic outcomes are qualitatively similar to those obtained for Poisson networks and vaccines that purely affect susceptibility, except for the fact that the homophily level at which the epidemic size is maximized is no longer independent of vaccine coverage. 

\paragraph{Conclusions and discussion.}

We have studied the effect of vaccination homophily, i.e., assortative mixing by vaccination status, on the herd immunity threshold and the expected epidemic size. In human society, vaccination homophily can emerge due to the presence of confounding factors, such as age~\cite{Mossong2008Social}, geography~\cite{Omer2008Geographic, Takahashi2017Geography}, socio-economic status~\cite{Danis2010Socioeconomic}, and personal and religious beliefs~\cite{Gastanaduy2016Measles}, that influence both the likelihood of interaction between individuals and the likelihood of them being in a common vaccination status. It can also occur as a consequence of behavioral contagion~\cite{Campbell2013Complex, Konstantinou2021Transmission} or inequality in the access to the vaccine. 
Our analysis is built on a model that embodies a minimalistic departure from the traditional assumption of homogeneous mixing and shows that the vaccination threshold for herd immunity is higher for stronger vaccination homophily. This suggests that herd immunity is more difficult, if not impossible, to achieve in the presence of vaccination homophily. It also implies that the well-known formula of Eq.~\eqref{eq:trad} underestimates the vaccination threshold by not taking homophily into account. 

We also show that the behavior of epidemic size as a function of homophily varies depending on the vaccine efficacy against susceptibility; when the efficacy is high, homophily monotonically amplifies the epidemic, while the epidemic size peaks at a nontrivial level of homophily when the efficacy is low. This is due to the competition between the herd immunity effect by homogeneous mixing and the epidemic containment by segregation. 
We can identify the parameter values for which homophily has a large impact on the epidemic size, which will have direct implications for the design of intervention strategies. 

Apart from vaccination homophily, another important type of inhomogeneity in networked epidemics is degree heterogeneity. Namely, real-world epidemics often exhibit a large variance in the number of secondary infections, whose distribution can be modeled by a negative binomial distribution~\cite{Lloyd-Smith2005Superspreading, Hebert-Dufresne2020Beyond}. The herd immunity threshold given by Eq.~\eqref{eq:critial_vax_rate_general} is not affected by the overdispersion of the distribution, but the epidemic size depends on the full shape of the distribution and therefore differs from the one for a Poisson network, as shown in the Supplemental Material. 
After the completion of this manuscript, we became aware of two other research works~\cite{Watanabe2021Impact, Burgio2021Homophily2} that report results in line with what we have described here. They found qualitatively similar effects of homophily on epidemic size for scale-free networks~\cite{Watanabe2021Impact} and empirical contact networks~\cite{Burgio2021Homophily2}. This further corroborates the generalizability of our theoretical findings to networks with heterogeneous degree distributions.

As a final remark, we note that our approach has a broader scope. In this Letter, we focused on homophily by vaccination status; however, our framework is general enough to account for homophily by adherence to other epidemic interventions that reduce the susceptibility or infectiousness of individuals, such as the practice of social distancing \cite{Sajjadi2021Social}, use of protective equipment~\cite{Watanabe2021Impact}, and adoption of digital contact tracing~\cite{Burgio2021Homophily, Rizi2021Epidemic}.
It can also be applied to the analysis of herd immunity in the case where the past infection (and consequent disease-induced immunity) is localized to a subpopulation~\cite{Britton2020Mathematical} and in the case where the mixing pattern is assortative by risk factors of the disease~\cite{Lemieux-Mellouki2016Assortative}.

\begin{acknowledgments}
\paragraph{Acknowledgements.}
MK acknowledges support from the project 105572 NordicMathCovid as part of the Nordic Programme on Health and Welfare funded by NordForsk.
The authors wish to acknowledge Aalto University ``Science-IT'' project for generous computational resources.
\end{acknowledgments}

\clearpage

\onecolumngrid
\begin{center}
\textbf{\large Supplemental Material}
\end{center}
\setcounter{equation}{0}
\setcounter{figure}{0}
\setcounter{table}{0}
\setcounter{page}{1}
\makeatletter
\renewcommand{\theequation}{S\arabic{equation}}
\renewcommand{\thefigure}{S\arabic{figure}}
\renewcommand{\bibnumfmt}[1]{[S#1]}
\renewcommand{\citenumfont}[1]{S#1}

\section{Network simulation}
The network simulation to obtain the giant component sizes is implemented in the following way.
First, we generate a network of size $N$ with vaccine coverage $\pi_\mathrm{v}$, strength of vaccination homophily $h$, and degree distribution $P(k)$. To do so, we begin by randomly assigning a vaccination status $\omega_i \in \{\mathrm{v}, \mathrm{u}\}$ and a degree $k_i$ to every node $i$ according to $\pi_\mathrm{v}$ and $P(k)$. Each of the $k_i$ stubs (half links) emanating from node $i$ is classified as an in-group stub with probability $\pi_{\omega_i \omega_i} = \pi_{\omega_i} + (1 - \pi_{\omega_i}) h$ and as an inter-group stub otherwise. Finally, we randomly make pairs of two in-group stubs within each group and pairs of an inter-group stub from each of the two groups, and connect the pairs to make a network. 

From this network, we randomly remove a $f_\mathrm{S}$ fraction of vaccinated nodes and compute the size of the giant component (defined as the largest connected component larger than 1\% of the unvaccinated population). The epidemic size is calculated as the mean giant component size over 20 network realizations. 

\section{Maximum effect of homophily on epidemic size}

The effect that vaccination homophily $h$ exerts on epidemic size $s$ varies for different parameter values. To characterize it, we compare the maximum size $s_\mathrm{max}$ of a large epidemic for homophily strengths in the range of $0 \leq h \leq 1$ with the baseline epidemic size $s_0$ under homogeneous mixing ($h=0$) for given values of $R_0$, $\pi_\mathrm{v}$, and $f_\mathrm{S}$. We fix $f_\mathrm{I} = 0$. 

\begin{figure}
    \centering
    \includegraphics[width=\textwidth]{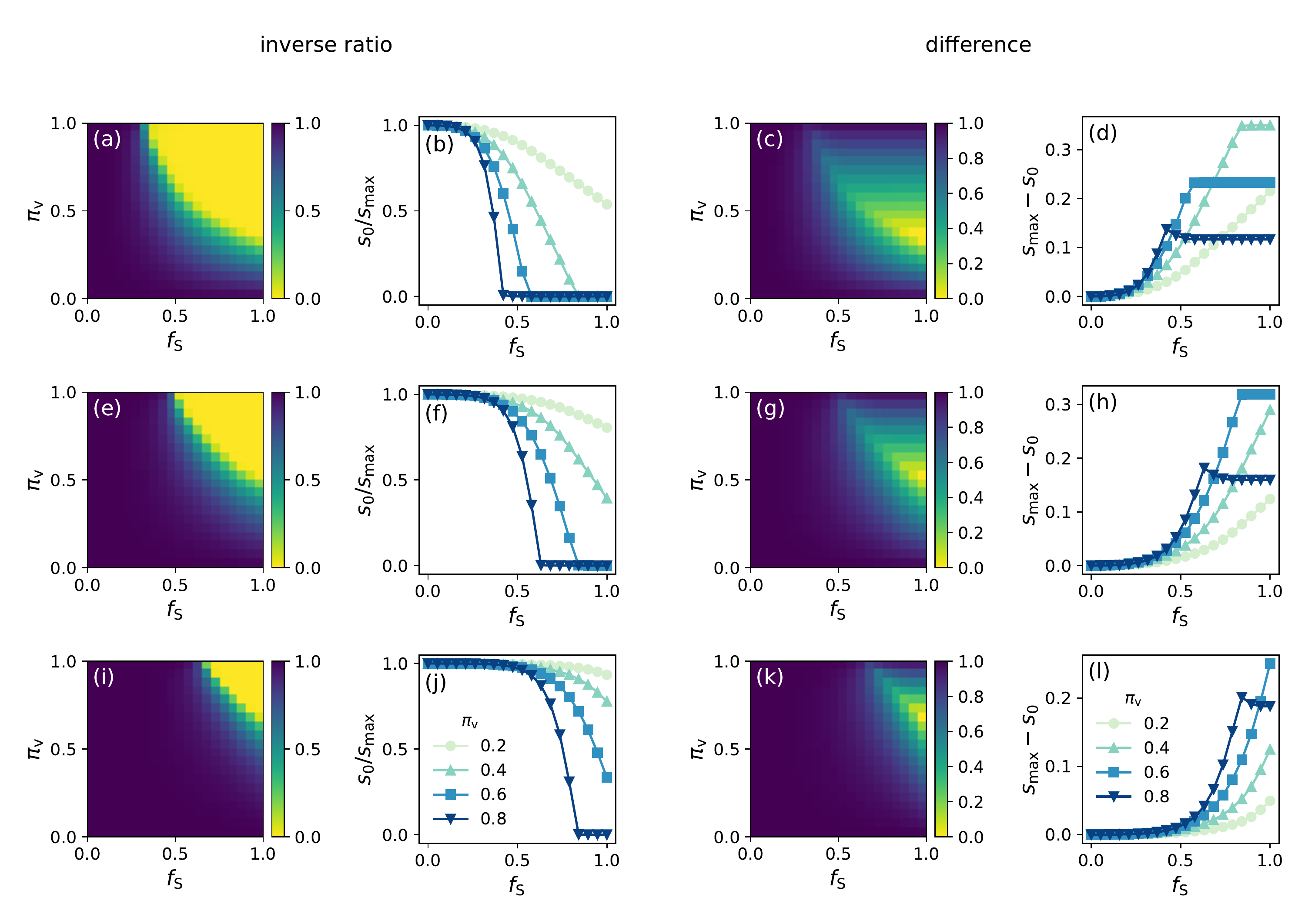}
    \caption{The maximum effect of homophily on epidemic size, measured by the inverse ratio $s_0 / s_\mathrm{max}$ (the left two columns) and the difference $s_\mathrm{max} - s_0$ (the right two columns), where $s_\mathrm{max}$ denotes the maximum epidemic size and $s_0$ is the baseline epidemic size $s_0$. Lighter colors represents larger discrepancies between the two epidemic sizes in the heatmaps. The basic reproduction numbers are: (a--d) $R_0 = 1.5$; (e--g) $R_0 = 2$; (i--l) $R_0 = 3$.}
    \label{fig:max_homophily_effect}
\end{figure}

The left two columns in Fig.~\ref{fig:max_homophily_effect} show the ratio $s_0/s_\mathrm{max}$ for different parameter values. The ratio takes a value close to one when either the vaccine coverage $\pi_\mathrm{v}$ or the efficacy $f_\mathrm{S}$ is small, indicating that the effect of homophily is small. In such a case, the herd immunity effect is already weak even under homogeneous mixing; therefore, its decline by homophily does not bring about a substantial difference in the epidemic outcome. In contrast, the ratio drops to zero when a highly effective vaccine covers a large fraction of the population. In this parameter region, the herd immunity is achieved at $h=0$ (i.e, $s_0 = 0$), but the presence of homophily brings the system out of the disease-free equilibrium. 

The ratio cannot describe how many additional infections will be caused by homophily, which is especially problematic when $s_0 = 0$. To this end, the difference $s_\mathrm{max} - s_0$ is calculated and shown in the right two columns in Fig.~\ref{fig:max_homophily_effect}. By comparing Fig.~~\ref{fig:max_homophily_effect}(a, e, i) and (c, g, k), we see that the difference between the maximum and baseline epidemic sizes is remarkably large around the herd immunity threshold at $h=0$ and, interestingly, is the largest when the vaccine is perfect.

% \begin{figure}
%     \centering
%     \includegraphics[width=0.5\textwidth]{max_homophily_inverse_ratio.pdf}
%     \caption{The ratio of the baseline epidemic size $s_0$ to the maximum epidemic size $s_\mathrm{max}$. The panels on the left column are heatmaps of the ratio $s_0 / s_\mathrm{max}$ as a function of $f_\mathrm{S}$ and $\pi_\mathrm{v}$. On the right column, we plot the ratio as a function of $f_\mathrm{S}$ for different values of $\pi_\mathrm{v}$. The basic reproduction numbers are: (a, b) $R_0 = 1.5$; (c, d) $R_0 = 2$; (e, f) $R_0 = 3$.}
%     \label{fig:max_homophily_effect_ratio}
% \end{figure}

% \begin{figure}
%     \centering
%     \includegraphics[width=0.5\textwidth]{max_homophily_diff.pdf}
%     \caption{The difference between the maximum epidemic size $s_\mathrm{max}$ and the baseline epidemic size $s_0$. The colors in the heatmaps on the left column denote the difference $s_\mathrm{max} - s_0$. The basic reproduction numbers are: (a, b) $R_0 = 1.5$; (c, d) $R_0 = 2$; (e, f) $R_0 = 3$.}
%     \label{fig:max_homophily_effect_diff}
% \end{figure}

\section{Imperfect vaccine against susceptibility and infectiousness}

\begin{figure}
    \centering
    \includegraphics[width=0.8\textwidth]{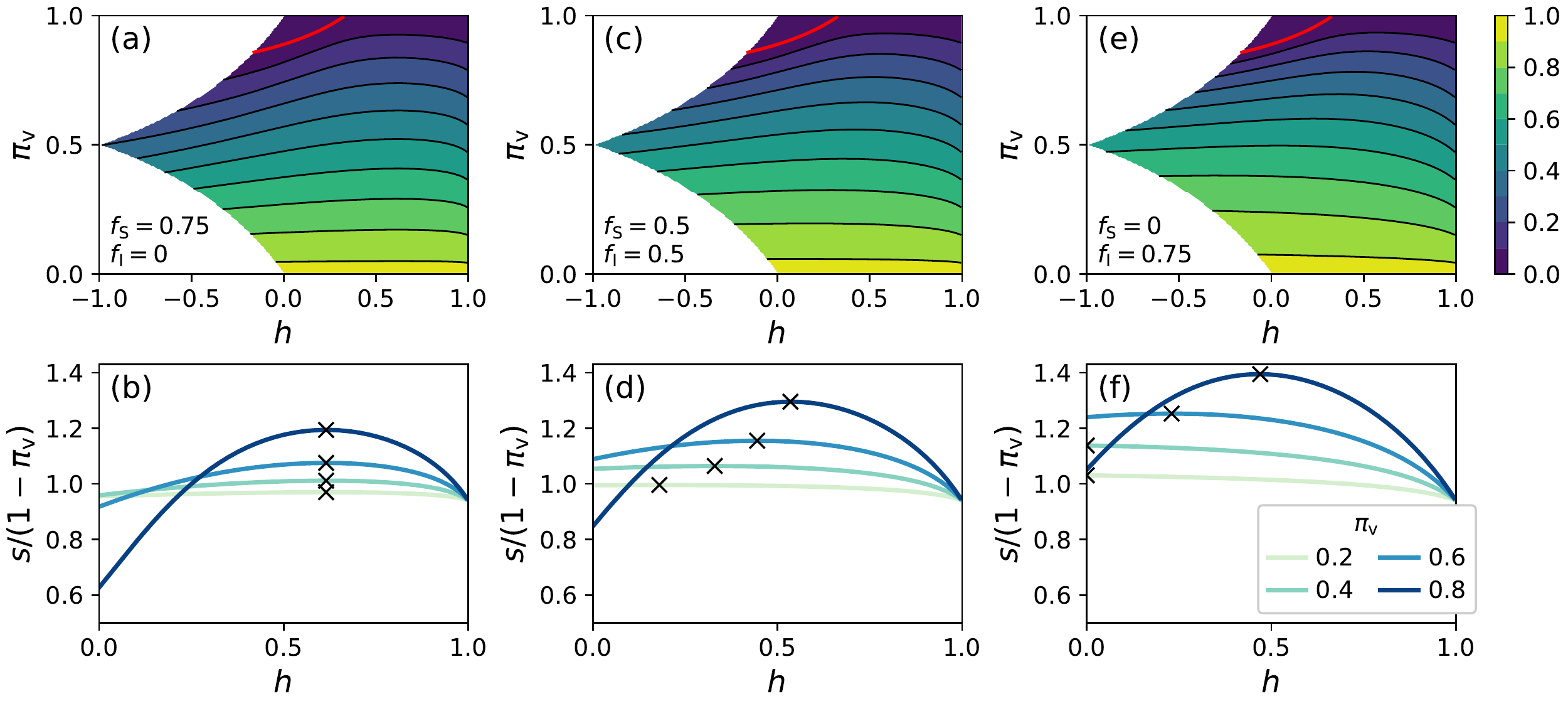}
    \caption{Epidemic size as a function of homophily strength $h$ and coverage $\pi_\mathrm{v}$ in Poisson networks under the coverage of imperfect vaccines. The vaccine is effective (a, b) only against susceptibility ($f_\mathrm{S} = 0.75, f_\mathrm{I} = 0$), (c, d) against both susceptibility and infectiousness ($f_\mathrm{S} = 0.5, f_\mathrm{I} = 0.5$), (e, f) only against infectiousness ($f_\mathrm{S} = 0, f_\mathrm{I} = 0.75$). Note that $\epsilon = (1 - f_\mathrm{S}) (1 - f_\mathrm{I}) = 0.25$ for all three vaccines, leading to the same herd immunity threshold (solid red line in the top panels). }
    \label{fig:episizes_poisson_eff_varied}
\end{figure}

\begin{figure}
    \centering
    \includegraphics[width=\textwidth]{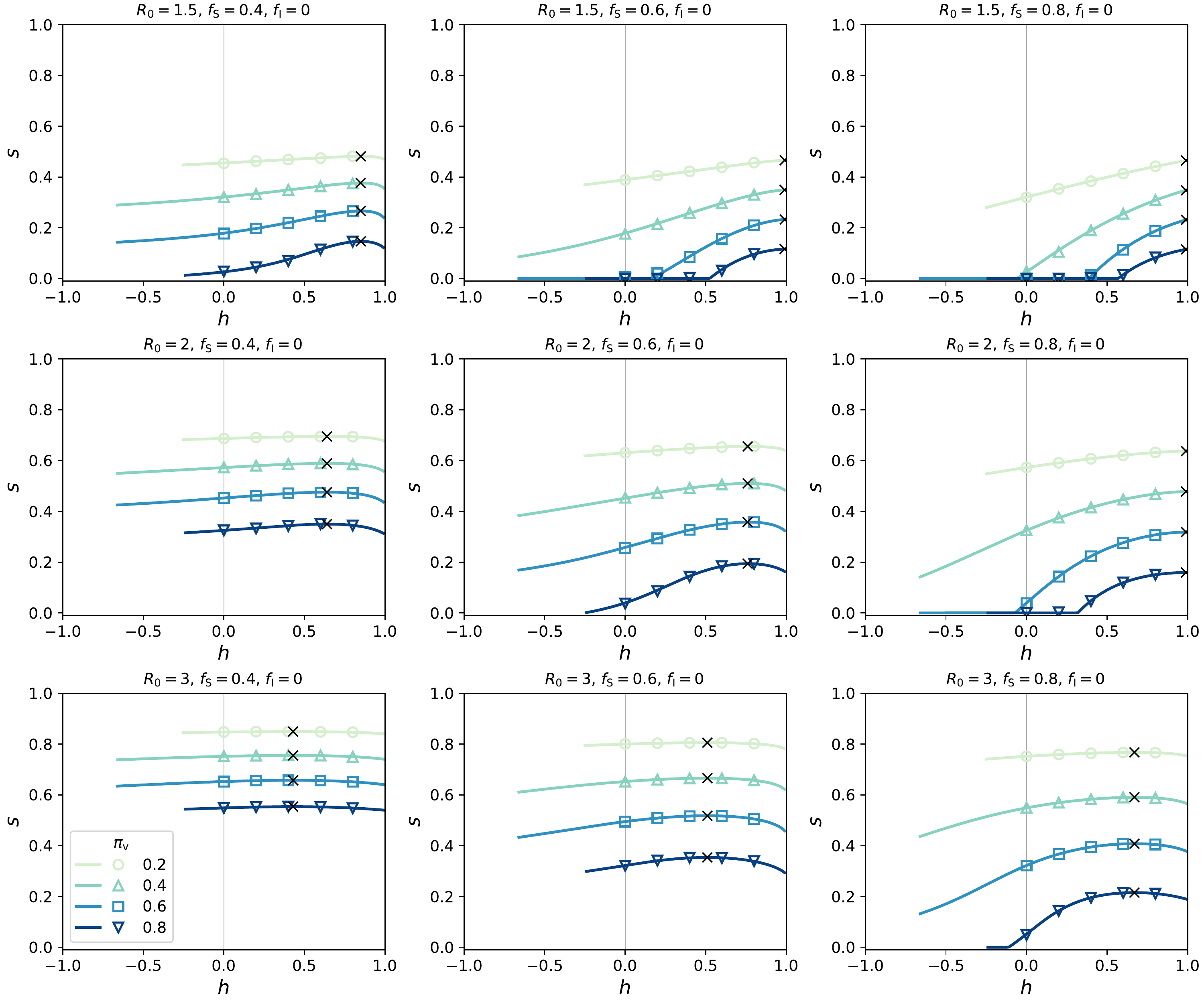}
    \caption{Epidemic size under the coverage $\pi_\mathrm{v}$ of vaccines that reduce susceptibility with efficacy $f_\mathrm{S}$. The cross symbols show the maximum point of each curve. The symbols denote epidemic sizes obtained by network simulation. Note that the curves do not extend to $h=-1$ because networks are unrealizable in the high heterophily region.}
    \label{fig:episizes_poisson_eff_sus_varied}
\end{figure}

\begin{figure}
    \centering
    \includegraphics[width=\textwidth]{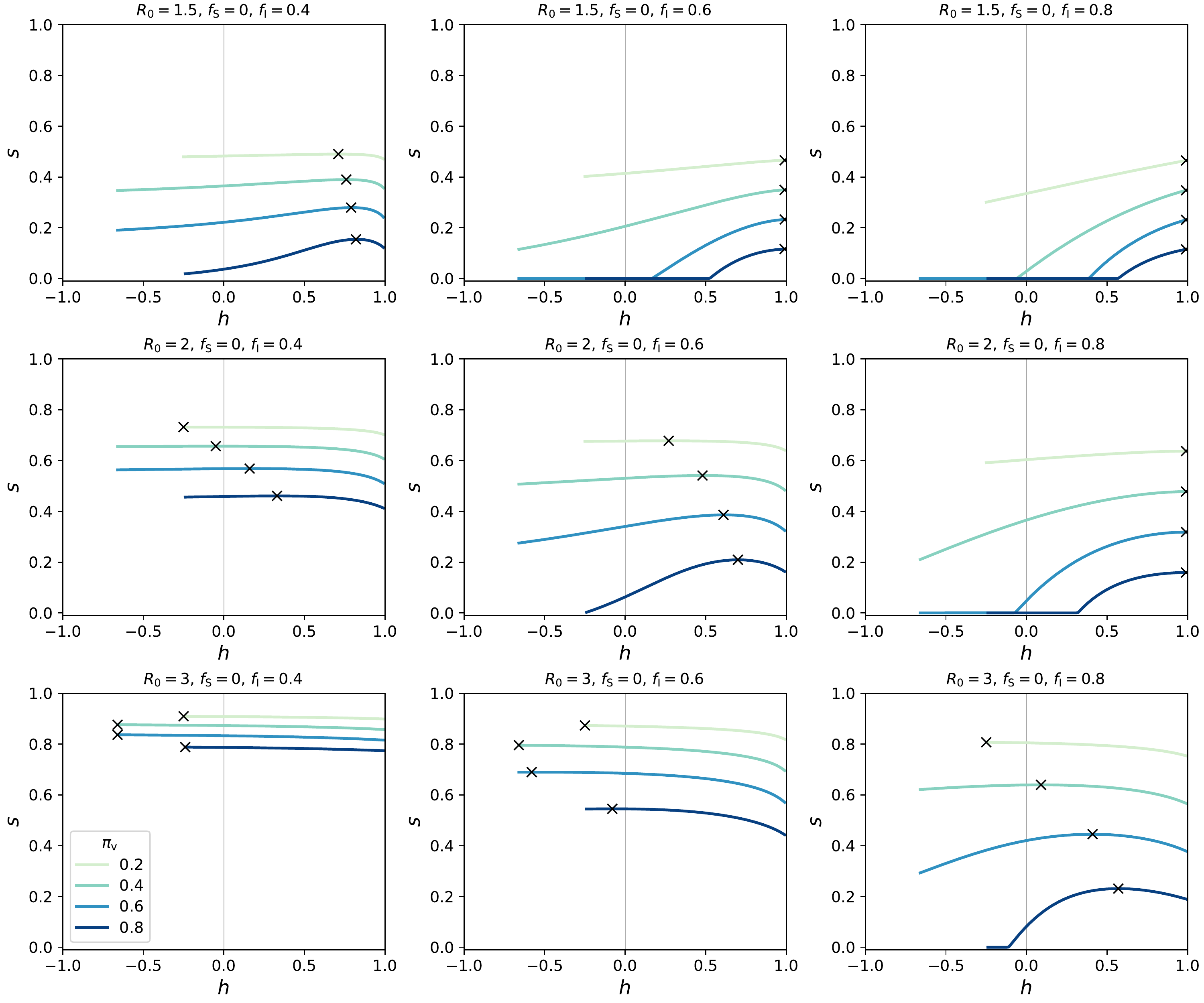}
    \caption{Epidemic size under the coverage $\pi_\mathrm{v}$ of vaccines that reduce infectiousness with efficacy $f_\mathrm{I}$. The cross symbols show the maximum point of each curve.}
    \label{fig:episizes_poisson_eff_inf_varied}
\end{figure}

In this section, we study the effect of vaccine efficacy against infectiousness. Figure~\ref{fig:episizes_poisson_eff_varied} shows the comparison between the epidemic sizes under the coverage of a vaccine purely against susceptibility, a vaccine against both susceptibility and infectiousness, and a vaccine purely against infectiousness. The value of $\epsilon$ is equal ($\epsilon = 0.25$) in all three cases; therefore, the herd immunity thresholds occur along the same line. For imperfect vaccines that reduce infectiousness, the homophily strength $h^*$ that maximizes the epidemic size is not independent of the vaccine coverage $\pi_\mathrm{v}$. Moreover, the maximum can occur at $h^* \leq 0$; that is, homophily makes the epidemic smaller for low vaccine coverage. 
Figures~\ref{fig:episizes_poisson_eff_sus_varied} and \ref{fig:episizes_poisson_eff_inf_varied} show the epidemic sizes for different combinations of the values of $R_0$, $\pi_\mathrm{v}$, and $h$ for imperfect vaccines that only reduce susceptibility and infectiousness, respectively.

\section{Negative binomial random network model}

In the main text, we focused on the network structure with the Poisson degree distribution. While the model represents the simplest way to integrate homophily with the canonical Erd\H{o}s-R\'{e}nyi random graph, it may be insufficient to describe the overdispersed spreading patterns seen in real-world epidemics~\cite{S_Lloyd-Smith2005Superspreading}. To add more realism, we consider the transmission networks in which the excess degree is negative-binomially distributed~\cite{S_Hebert-Dufresne2020Beyond} with mean $R_0$ and the dispersion parameter $r$ as
\begin{equation}
    \tilde{P}(\tilde{k}) = \binom{\tilde{k} + r - 1}{\tilde{k}} \left(\frac{r}{R_0 + r}\right)^{r} \left(\frac{R_0}{R_0 + r}\right)^{\tilde{k}},
\end{equation}
where we denote the excess degree and its distribution by tildes. This distribution converges to the Poisson distribution with mean $R_0$ in the limit $r \to \infty$; hence the dispersion $r$ quantifies the deviation from the Poisson. In the following, we derive the degree distribution $P(k)$ consistent with this excess degree distribution by considering its probability generating functions (PGF). This derivation is based on H\'{e}bert-Dufresne~et~al.~\cite{S_Hebert-Dufresne2020Beyond}. 

In general, the PGF of the degree $g_0$ and that of the excess degree $g_1$ are related as follows: 
\begin{equation}
    g_1(x) = \frac{g_0'(x)}{g_0'(1)}\,,
\end{equation}
where $g_0'(x)$ denotes the derivative of $g_0(x)$ with respect to $x$. This implies that $g_1(x)$ constrains $g_0(x)$ up to a constant and a multiplicative factor. Namely, we have
\begin{equation}
    g_0(x) = g_0'(1) \int g_1(x) dx,
    \label{eq:g_0_from_g_1}
\end{equation}
with boundary conditions $g_0(0) = p_0$ and $g_0(1) = 1$. Here, $p_0$ is a free parameter satisfying $0 \leq p_0 \leq 1$.

The PGF of the negative binomial distribution for the excess degree is given by
\begin{equation}
    g_1(x) = \left[1 + \frac{R_0 (1 - x)}{r}\right]^{-r}.
\end{equation}
Plugging this into Eq.~\eqref{eq:g_0_from_g_1}, we obtain
\begin{equation}
    g_0(x) = p_0 + (1 - p_0) \left[1 - \left(\frac{r}{R_0 + r}\right)^{1 - r}\right]^{-1} \left[1 - \left(1 - \frac{R_0 x}{R_0 + r}\right)^{1 - r}\right]
\end{equation}
for $r \neq 1$ and 
\begin{equation}
    g_0(x) = p_0 + (1 - p_0) \left[1 - \frac{\log [1 + R_0 (1 - x)]}{\log [1 + R_0]} \right]
\end{equation}
for $r = 1$. 
By taking the $k$-th derivative of $g_0$ as $g_0^{(k)}(x) = d^k g_0(x) / dx^k$, we recover the degree distribution:
\begin{equation}
    P(k) = \frac{g_0^{(k)}(0)}{k!} = \begin{cases}
    p_0, & \text{(for } k = 0 \text{)}\\
    - (1 - p_0) \dbinom{k + r - 2}{k} \left[1 - \left(\dfrac{r}{R_0 + r}\right)^{1 - r}\right]^{-1}  \left(\dfrac{R_0}{R_0 + r}\right)^{k} & \text{(for } k > 0, r \neq 1 \text{)}\\
    \dfrac{(1 - p_0)}{k \log (1 + R_0)}\left(\dfrac{R_0}{1 + R_0}\right)^k & \text{(for } k > 0, r = 1 \text{)}.
    \end{cases}
\end{equation}
If $r > 1$, setting $p_0 = [r / (R_0 + r)]^{r-1}$ makes the degree distribution reduce to a negative binomial distribution
\begin{equation}
    P(k) = \binom{k + r - 2}{k} \left(\frac{r}{R_0 + r}\right)^{r - 1} \left(\frac{R_0}{R_0 + r}\right)^{k}.
\end{equation}
The mean and dispersion parameters of this distribution are equal to $(r - 1)R_0 / k$ and $r - 1$, respectively.

The fraction of isolated nodes $p_0$ effectively controls the system size because these nodes will not be part of the epidemic. As our motivation to introduce the negative binomial distribution is to see the effect of overdispersion in excess degree as compared to the Poisson distribution at the limit of $r \to \infty$, we do not want the value of $p_0$ to affect the comparison. A natural choice would then be to set the value of $p_0$ to be equal to the fraction of isolated nodes for the corresponding Poisson degree distribution, that is, $p_0 = e^{-R_0}$.

\begin{figure}
    \centering
    \includegraphics[width=0.8\textwidth]{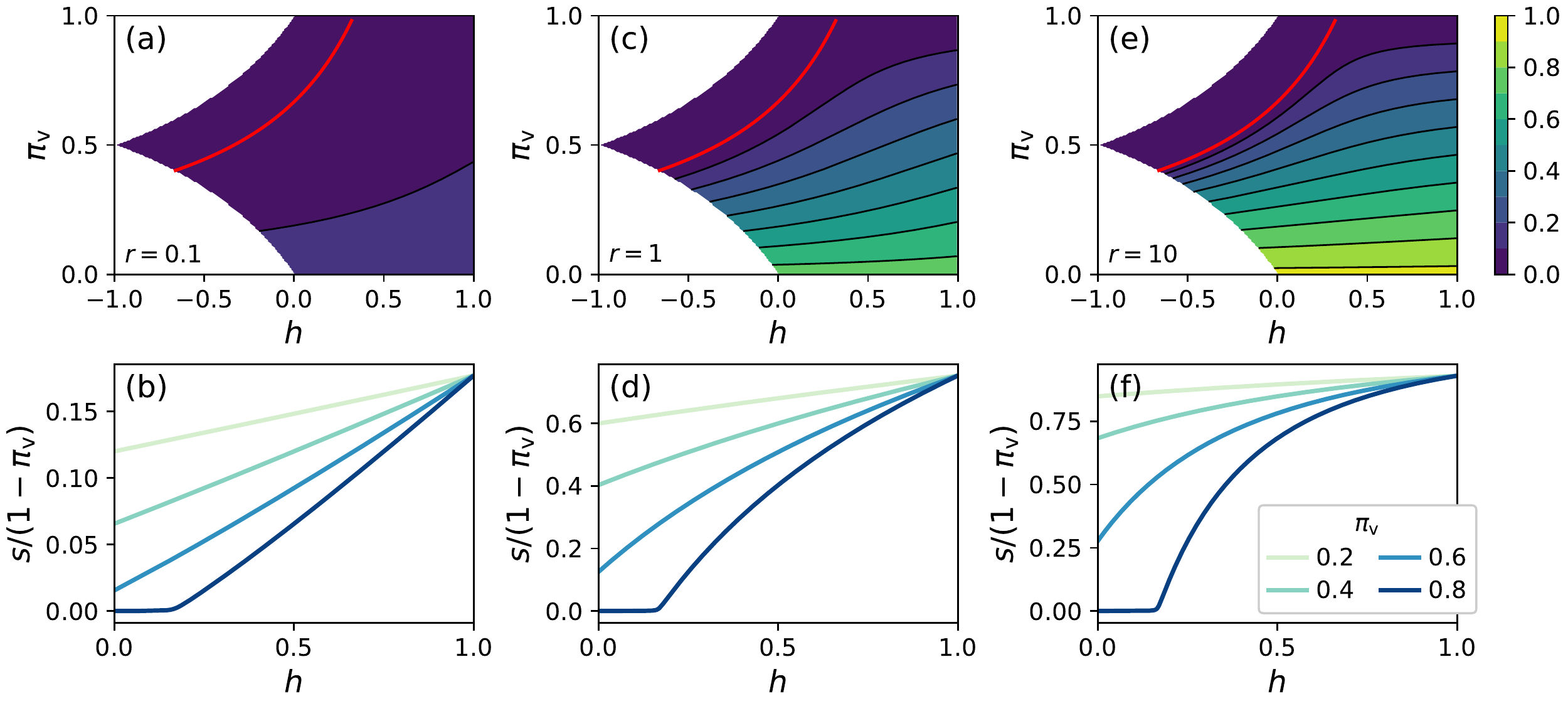}
    \caption{Epidemic size as a function of homophily strength $h$ and coverage $\pi_\mathrm{v}$ in negative binomial networks under the coverage of a perfect vaccine. Only the dispersion parameters are varied: (a, b) $r=0.1$; (c, d) $r=1$; (e, f) $r=10$. The basic reproduction number $R_0 = 3$ in all panels.}
    \label{fig:episizes_nbinom_disp_varied}
\end{figure}

The epidemic size under the coverage of a perfect vaccine in negative binomial networks with $R_0 = 3$ and different values of the dispersion parameter $r$ are shown in Fig.~\ref{fig:episizes_nbinom_disp_varied}. When $r=10$, the epidemic size is similar to that for a Poisson network. Decreasing the dispersion parameter (i.e., increasing the variance) does not change the herd immunity threshold but decreases the epidemic size when the system is in the epidemic state. For example, if the system is maximally homophilic, i.e., $h=1$, so that the vaccinated and unvaccinated subpopulations are completely isolated from each other, 93\% of the unvaccinated will get infected when the variance in excess degree is relatively small with $r=10$; however, this fraction decreases to 75\% for $r=1$ and to 18\% for $r=0.1$. Moreover, one can see from the bottom row of Fig.~\ref{fig:episizes_nbinom_disp_varied} that the critical behavior as a function of $h$ changes when the value of $r$ is varied. 

\begin{figure}
    \centering
    \includegraphics[width=0.8\textwidth]{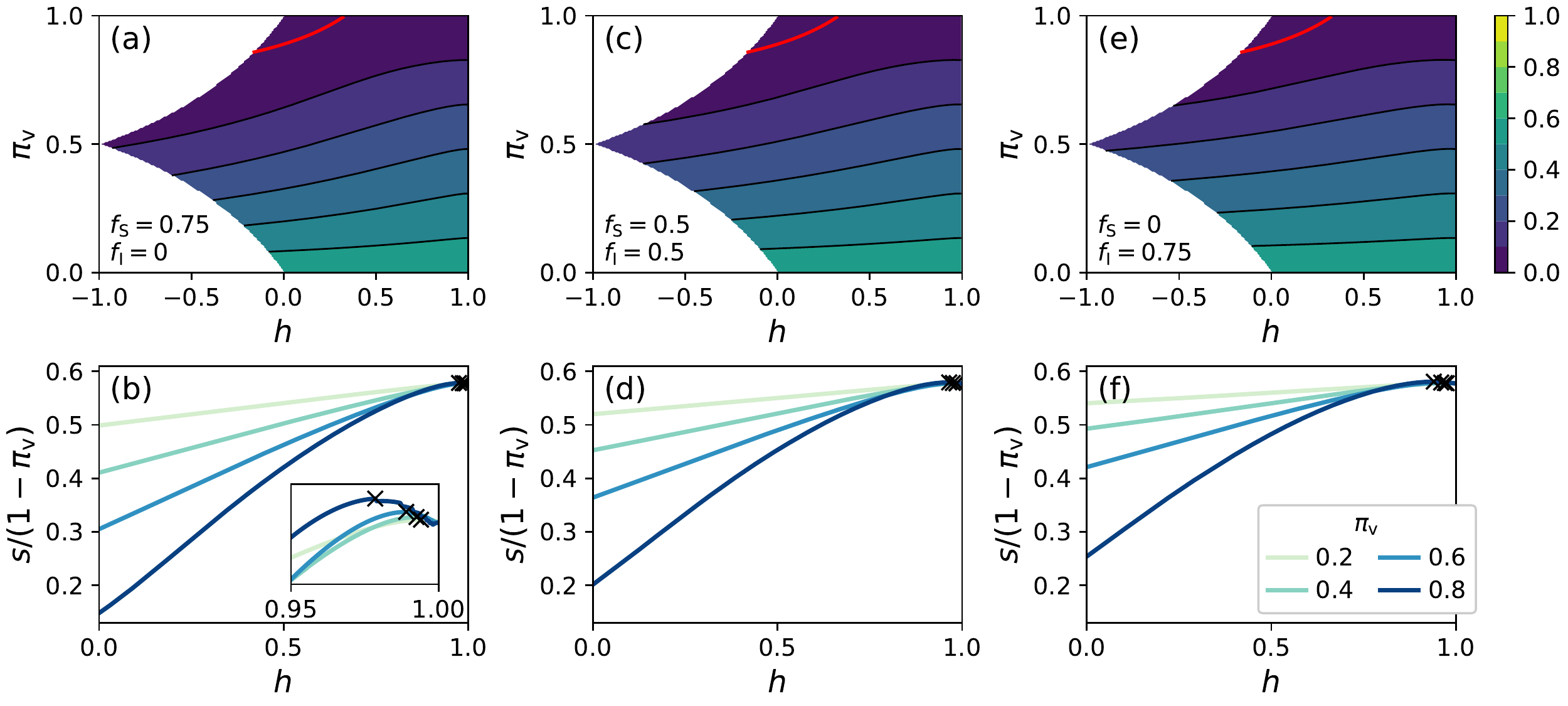}
    \caption{Epidemic size as a function of homophily strength $h$ and coverage $\pi_\mathrm{v}$ in negative binomial networks under the coverage of imperfect vaccines. The vaccine efficacies are the same as in Fig.~\ref{fig:episizes_poisson_eff_sus_varied}, that is, (a, b) only against susceptibility ($f_\mathrm{S} = 0.75, f_\mathrm{I} = 0$), (c, d) against both susceptibility and infectiousness ($f_\mathrm{S} = 0.5, f_\mathrm{I} = 0.5$), (e, f) only against infectiousness ($f_\mathrm{S} = 0, f_\mathrm{I} = 0.75$). The basic reproduction number $R_0 = 3$ and dispersion parameter $r = 0.5$ in all panels.}
    \label{fig:episizes_nbinom_eff_varied}
\end{figure}

The overdispersed degrees of negative binomial networks affect the epidemic size under the coverage of imperfect vaccines too. A marked qualitative difference from the Poisson case is that the homophily strength $h^*$ maximizing the epidemic size is not independent of the vaccine coverage even for vaccines purely against susceptibility (Fig.~\ref{fig:episizes_nbinom_eff_varied}(b)). Another observation is that, compared to the Poisson case (Fig.~\ref{fig:episizes_poisson_eff_varied}), the epidemic size is maximized at $h$ close to one. In other words, the amplification of epidemic size by homophily is more robustly observed in overdispersed networks.

\end{document}